\documentclass[aip,jcp,reprint]{revtex4-1}
\usepackage{graphicx} %eps figures can be used instead
\usepackage[format=plain,justification=raggedright,singlelinecheck=false,font=small,labelfont=bf,labelsep=space]{caption}
\usepackage{color}

\begin{document}

\title{Aging near rough and smooth boundaries in colloidal glasses}
\author{Cong Cao}
\email{ccao8@emory.edu}
\author{Xinru Huang}
\author{Connie B.~Roth}
\author{Eric R.~Weeks}
\email{erweeks@emory.edu}
\affiliation{Department of Physics, Emory University, Atlanta, GA 30322, USA}
\date{\today}

\begin{abstract}
We use confocal microscopy to study the aging of a bidisperse
colloidal glass near rough and smooth boundaries.  Near smooth
boundaries, the particles form layers, and particle motion is
dramatically slower near the boundary as compared to the bulk.
Near rough boundaries, the layers nearly vanish,
and particle motion is nearly identical to that of the bulk.
The gradient in dynamics near the boundaries is demonstrated to
be a function of the gradient in structure for both types of
boundaries.
Our observations show that wall-induced layer structures strongly
influence aging.
\end{abstract}

% \pacs{not yet}

\maketitle

\section{Introduction}
\label{Introduction}

Glasses are solids with disordered structures
and slow internal dynamics.  Efforts to
understand the influence of boundaries on glassy
dynamics has been an active area of research for more than two
decades.\cite{alcoutlabi05,forrest01,richert11,roth05,baschnagel05,ediger14,napolitano17,hunter12rpp}
Initial efforts on confined systems
were thought to provide a route to accessing
postulated growing length scales associated with cooperative
motion.\cite{baschnagel05,binder03,kob00,scheidler02,pye10,hanakata14,lang13,sarangapani11,sarangapani12}
However, the study of such small system sizes necessitates
the presence of boundaries and it has turned out that the
specific details of such interfaces have a great deal of
influence on the local dynamics near the boundary.\cite{he07}
In experimental material systems, the type of interface often
plays a dominant role over finite size effects where interfacial
energy, specific chemical interactions, and substrate compliance
are all factors that have shown to have some influence on the
dynamics.\cite{fryer01,tsui01,paeng12,priestley05,roth07,evans12,baglay15,wang13,christie16,edmond12,hunter14,baglay17}
In computer molecular dynamics (MD) simulations where the specific
details of the boundary need to be constructed at its most basic
level, it is unclear a priori how best to accomplish this.

Early MD efforts started with smooth, structureless walls
where the boundary was treated as a continuum and details
of the wall potential were integrated over in the lateral
($x$,$y$) direction leaving only a $z$-dependence perpendicular
to the boundary.\cite{baschnagel05}  Alternatively, molecularly
structured walls assembled from Lenard-Jones (LJ) particles into
either crystalline arrays or frozen amorphous structures were also
investigated.\cite{kob00,scheidler02,scheidler04,hanakata15,davris15,varnik02}
In these simulations, local dynamic near the boundary were
usually different than bulk, but the underlying cause why
was frequently unclear.  Smooth walls typically exhibit
faster dynamics than bulk in part because there is no
penalty for the particles to slide laterally along the
wall,\cite{binder03,varnik02e,baschnagel02,peter06,hanakata12}
a type of motion only considered to be experimentally
relevant for a free surface.\cite{peter06}  Systems with
molecularly structured walls, where lateral sliding
is inhibited, typically exhibited slower dynamics in
comparison.\cite{kob00,scheidler02,hanakata14,scheidler04,hanakata15,smith03}

One of the major challenges with such boundaries is that for
mixtures of LJ particles or polymeric bead-spring models (the most
commonly modeled systems), the presence of the wall creates layering
of the particle density $\rho(z)$ as a function of distance from the
wall.\cite{baschnagel05}  Intuitively, the particles pack easily
in a layer against the wall, and then the particles in the second
layer pack against that first layer, {\it etc.}, with the influence
of the wall diminishing farther away.  Thus, a major effort in these
studies is the need to determine the extent to which the observed
differences in local dynamics a distance $z$ to the boundary are
influenced by the local $\rho(z)$ structure in density.  In some
cases slower dynamics near the boundary has been associated with a
significantly increased local density,\cite{baschnagel05,fehr95}
while other studies have demonstrated that the change in
dynamics near the boundary is unrelated to the $\rho(z)$ density
profile.\cite{kob00,scheidler02,hanakata15,hanakata12} For example,
even efforts to construct a ‘neutral’ boundary that avoids local
perturbations to the particle density by freezing in an amorphous,
liquid-like structure still leads to perturbations in the local
dynamics.\cite{kob00,scheidler02,scheidler04,biroli08,li14,hocky14,himanagamanasa15}
It is important to note that local perturbations to the $\rho(z)$
structure are not limited to only coarse-grained simulations,
they are also observed in nearly-atomistic, united-atom
models.\cite{hudzinskyy11}  In addition, experimental
studies on glassy thin film systems are also trying to
uncover the extent to which  molecular ordering occurs near a
boundary and its possible influence on the local density and
dynamics.\cite{gin12,sen16,napolitano13,gautam00,huang16}

Here we present a direct experimental comparison of local glassy
dynamics next to rough and smooth boundaries using colloidal
glasses, which have been previously suggested as a means of
experimentally verifying these observations from coarse-grained
MD simulations of boundaries.\cite{hunter12rpp,fehr95,nemeth99}
Colloids are small solid particles in
a liquid, where Brownian motion allows particles to diffuse and
rearrange.\cite{hunter12rpp}
We use confocal microscopy to study the aging of a bidisperse
colloidal glass where layer-resolved dynamics as a function of
distance from a rough or smooth wall are compared with the measured
$\rho(z)$ density profile.  Smooth boundaries are simply a normal
untreated glass coverslip, while rough boundaries are constructed by
melting a small amount of the colloidal sample to the coverslip.
These stuck particles cover approximately 30-50\% of the surface and 
provide a roughness scale comparable to
the particle size.  The particle-glass and particle-particle
interactions are purely repulsive and so the main difference in
the boundary conditions is the topography.  We observe distinctly
different results between smooth and rough boundary conditions:
near smooth boundaries motion is dramatically slower, whereas
near rough boundaries the aging process is nearly independent of
the distance from the boundary.  We ascribe this to the strong
influence of layer-like structures formed near the smooth boundary.

Our samples are aging:  unlike many phases of matter, glasses
are out of equilibrium, and so their properties slowly evolve,
perhaps toward a steady
state.\cite{struik77,vanmegen98,priestley09,ramirez10aging}
These properties can include the density, enthalpy, 
and diffusive motion of the molecules comprising the glass.
This has implications for the usage of glassy materials which
have properties that depend on age perhaps in an undesirable
way.\cite{tant81,huang04polymer,priestley09}  Aging has been observed
in polymer glasses,\cite{priestley09,tant81} granular
systems,\cite{knight95,richard05} and soft materials such as colloids and
foams.\cite{vanmegen98,elmasri05,martinez08,cipelletti02,courtland03,lynch08,bandyopadhyay04,mckenna09,di11}
While for polymer glasses and granular materials aging is often
measured as slight decreases in volume, colloidal glasses are
typically studied at constant volume.  The main signature of
aging of colloidal glasses is the dramatic slowing of particle
motion as the sample ages,\cite{vanmegen98,elmasri05,martinez08}
often characterized by the slowing down of the mean square
particle displacement for time windows at increasing aging
times.\cite{courtland03} Previous work suggests that aging
in colloidal systems may relate to the local structure around
rearranging particles \cite{lynch08} or domains of more mobile
particles.\cite{courtland03,yunker09}  In general, it is not surprising
that confined glasses age in different ways from their bulk
counterparts.\cite{priestley09}   In this manuscript we show that
aging of colloidal particles is tied to layering structure imposed
by the nearby sample boundaries.

\section{Experimental Details}
\label{sectionMethod}

%particles are ASM75 and GK003

In our experiment we use sterically stabilized poly(methyl
methacrylate) (PMMA) particles
\cite{antl86,pusey86} to prevent aggregation. Two different sizes
of particles are mixed in order to prevent crystallization,
with $d_{\rm L} = 2.52$~$\mu$m and $d_{\rm S} = 1.60$~$\mu$m.
The particles have a polydispersity of 7\%.  The number ratio
is approximately $1:1$.  To match the particles' density and
refraction, we use a mixture of decalin and cyclohexylbromide as
the solvent.\cite{dinsmore01}   We view our samples with a
fast confocal microscope (VT-Eye from Visitech, International).
The large particles are dyed with rhodamine dye and thus are
visible, while the small particles are undyed and thus unseen.
Based on prior work, we expect that both small and large particles
have similar behavior \cite{lynch08,narumi11}.  Visual inspection
using differential interference contrast microscopy, which can
see both particle types, confirms that the particles are
well-mixed even at the boundary.
The imaged volume is
$50 \times 50 \times 20$ $\mu$m$^3$.  These images are taken
once per minute for 2 hours.  Our scanning volume starts about
5 $\mu$m outside the boundary to ensure we have clear
images of the particles at the boundary.  The microscope pixel
size is $0.11$~$\mu$m in $x$ and $y$ (parallel to the boundary)
and $0.2$~$\mu$m in $z$ (perpendicular to the boundary).  We use
standard software to track the motion of the particles in
3D.\cite{crocker96,dinsmore01}  Our particle locations
are accurate to 0.10~$\mu$m initially and to 0.15~$\mu$m by the
end of the experiment after some photobleaching occurs.

We construct two types of sample chambers.  The first uses a normal
untreated coverslip as a smooth boundary.  The second is prepared
by taking a small amount of colloidal sample and melting this
on to the coverslip, using an oven at 180$^\circ$~C for 20 min.
After this process, the PMMA particles are
irreversibly attached to the coverslip.  
This sample is the same bidisperse mixture of PMMA
particles as the main sample with the exception that both particle
sizes are undyed.  
By image analysis we
determine that the stuck particles cover approximately 30-50\%
of the surface.  The specific fraction is difficult to measure
as we only image the large fluorescently dyed particles, so we
cannot see either the smaller mobile particles of our bidisperse
sample or the stuck particles of either size.  After adding the
samples, we never observe any of our sample particles stuck to
the boundaries for either smooth or rough boundary conditions.

We add a stir bar inside each sample chamber so that we
can shear rejuvenate the samples \cite{courtland03} and thus
initiate the aging process and set $t_{\rm age}=0$ (when we stop
stirring).  Note that $t_{\rm age}$ is set by the laboratory
clock and thus is identical throughout our sample; we are not
considering the idea of a spatially varying time scale.
We find the stirring method gives reproducible results similar to
prior work,\cite{courtland03,lynch08} although this is probably
different from a temperature or density quench as is usually done
for polymer and small molecule glasses.\cite{mckenna03} Given the
flows caused by the stirring take 20-30~s to appreciably decay after
stirring is stopped, there is some uncertainty in our $t_{\rm age}
= 0$, but we examine the data on time scales at least ten times
larger than any uncertainty of this initial time.

Confocal microscopy allows us to measure the bulk number density
for large particles, which we find to be $0.32 \pm 0.03$~$\mu$m.
The uncertainty represents the variability that we see from
location to location.  Given that we cannot directly observe the
small particles, the observed number density in any given location
is not a useful measure of the local volume fraction.  Thus, we
do not have a direct measure of the volume fraction.  Given that
the samples behave as glasses (to be shown below) and particles are
still able to move, we conclude $\phi_{\rm glass} < \phi < \phi_{\rm
rcp}$, with $\phi_{\rm glass} \approx 0.58$ (for bidisperse colloidal
glasses\cite{narumi11}) and $\phi_{\rm rcp}
\approx 0.65$ (the value for random close packing for our
bidisperse sample\cite{desmond14}).  Prior studies
of aging colloidal glasses found little \cite{martinez10} or no
\cite{courtland03} dependence of the behavior on $\phi$.
It is important to note that we cannot definitely establish if our two
samples are at the same $\phi$ or, if not, which one would be
higher.

\section{Results}
\label{sectionResults}

\begin{figure}
\centering
\includegraphics[width=8cm,keepaspectratio]{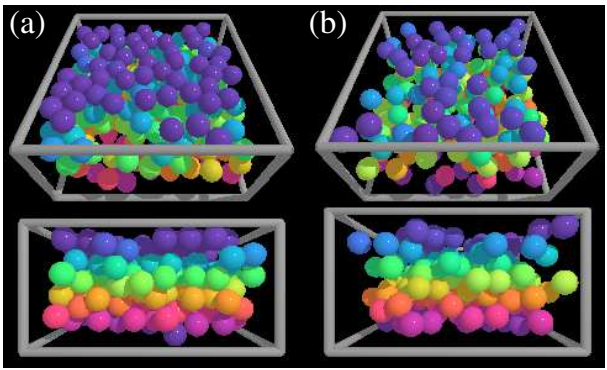}
\caption{Top view and side view for reconstructed $3D$ images
for colloidal samples near (a) a smooth boundary and (b) a rough
boundary.  Color is a continuous parameter representing particles’ distances
away from the boundary (from 0 to 10~$\mu$m). The particles
closest to the boundary are on the top and colored dark purple.
The grey boxes have dimensions $20 \times 20 \times 15$~$\mu$m$^3$,
which is a subset of the full image volume.
While the sample has particles of two sizes, only the large
particles are visible in the experiment.  The data are pictured
at $t_{\rm age}=10$~min.
}
\label{povray} 
\end{figure}

Figure \ref{povray} shows reconstructed 3D images for
smooth (a) and rough (b) boundaries.  To show the influence
of the boundaries, the particles closest to the boundary
are on the top of these pictures (colored dark purple).
The color changes continuously as a function of the distance $z$
away from the boundary.  However,
the particles shown in Fig.~\ref{povray}(a) appear to have
discrete colors as they form layers with distinct $z$ values.
This phenomena is induced by the flat wall and is well
known.\cite{landry03,edmond12}

\begin{figure}
\centering
\includegraphics[width=8cm,keepaspectratio]{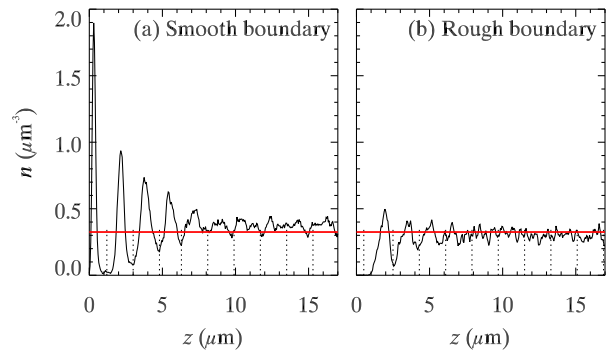}
\caption{The local number density $n(z)$ as a function of the
distance from the boundary at $z=0$~$\mu$m for samples near (a)
a smooth boundary and (b) a rough boundary.  Layer-like structures
are observed in both samples in first few layers, although they are
sharper for the smooth boundary and persist to larger $z$.
The vertical dotted lines indicate separate layers, with a fixed
spacing once the layers become ill-defined.  The red horizontal
lines show the average number density in the region $z >
10$~$\mu$m.
}
\label{numdensity} 
\end{figure}

To quantify the layered structure we measure the time-averaged
number density for the large particles $n(z)$.  This is shown in
Fig.~\ref{numdensity} for smooth (a) and rough (b) boundaries.
We set $z=0$ at the center of the particle whose center is
closest to the boundary.  The vertical dotted lines indicate
the separate layers.  As the
sample is composed of two sizes of particles, the layer structure
decreases rapidly away from the wall, consistent with simulations
\cite{kjellander91,mittal08} and experiments.\cite{edmond12} The
first peak in Fig.~\ref{numdensity}(a) has the maximum value and
minimum width, indicating particles are in a well-defined layer,
consistent with Fig.~\ref{povray}(a).  By the sixth layer, it
is unclear if there is still a layer or if we are seeing random
number density fluctuations.  For the rough wall in panel (b),
the layers become poorly defined by the fourth layer.  For later
analysis, we continue counting the layers by defining them in the
bulk region to be every 1.8~$\mu$m based on the typical spacing
of the well-defined layers.  Note that for the rough boundary
condition, the wall texture occupies some of the space of the
first layer, thus decreasing the number of dark purple particles
in Fig.~\ref{povray}(b) and reducing the area under the first peak
in Fig.~\ref{numdensity}(b).

\begin{figure}
\centering
\includegraphics[width=8cm,keepaspectratio]{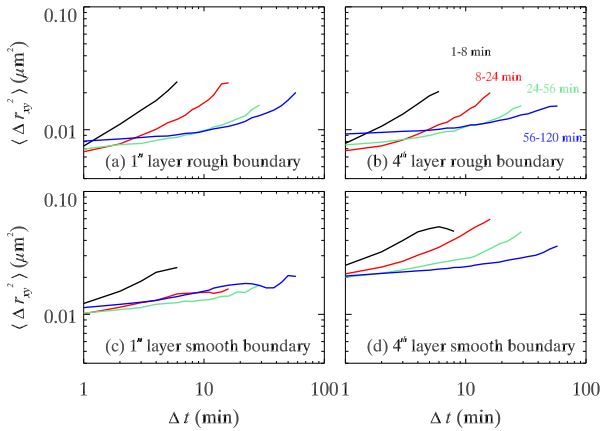}
\caption{The mean square displacement for motion
parallel to the boundaries calculated as $\Delta r_{xy}^2 = \frac{1}{2}
\langle \Delta x^2 + \Delta y^2 \rangle$.  The data are averaged
over four different $t_{\rm age}$ regimes as indicated.
Data are for (a)
$1^{st}$ layer with a rough boundary, (b) $4^{th}$ layer with
a rough boundary, (c) $1^{st}$  layer with a smooth boundary,
(d) $4^{th}$ layer with a smooth boundary.  The data for the
$4^{th}$ layers match the bulk behavior, and their progression to
larger time scales with increasing $t_{\rm age}$ demonstrates that the
sample is aging.  The data for the $1^{st}$ layers show that
aging is fairly unchanged for the rough boundary (a), but
markedly different for the smooth boundary (c).
All displacements are normalized by the
large particle diameter $d_L$.
}
\label{msdxy} 
\end{figure}

Aging manifests as a slow change of sample behavior with
increasing $t_{\rm age}$, where the rate of change slows at
longer times.\cite{struik77} The easiest quantity to see this
with our data is the mean square displacement (MSD) of particle
motion.\cite{lynch08,yunker09}  Fig.~\ref{msdxy} shows the motion
parallel to the boundaries for (a,b) rough and (c,d) smooth
boundaries, with panels (b) and (d) corresponding to the bulk MSD
curves.  The different colors indicate different ages.  The mean
square displacement is computed as $\frac{1}{2} \langle \Delta x^2 +
\Delta y^2 \rangle$ where the angle brackets indicate an average
over all large particles and over all starting times within the
window of $t_{\rm age}$.  For our shortest time scale ($\Delta t =
1$~min) the MSD curves have a shallow slope indicating particles are
trapped by the local configuration, with the exception of the black
curves ($t_{\rm age} \leq 8$~min) when the aging has just started.
At long time scales, the MSDs show an upturn, which is related to
the samples' age.\cite{courtland03,lynch08,yunker09}  For larger
$t_{\rm age}$ the lag time particles need to reach the same MSD
increases, indicating the slowing particle motion.  Note that
as we take data, the fluorescent dye in the particles begins to
photobleach and our particle tracking resolution worsens, slightly
increasing the measured MSD values at small $\Delta t$.\cite{poon12}
Slight differences in image quality may also be affecting the
overall height of the MSD curves between the smooth and rough
boundary conditions for the data at $\Delta t \leq
10$~min.\cite{poon12} Accordingly, for subsequent analysis below,
we will focus on large $\Delta t$ values for which the signal is
greater than the photobleaching noise.  The main points to learn
from Fig.~\ref{msdxy} are that the overall behavior of the curves
shows the expected aging trend with larger $t_{\rm age}$, and
panels (b,d) show the aging curves are similar for both boundary
conditions far from the boundary.

Figure \ref{msdxy}(a) shows the MSD curves for $xy$ motion for the
first layer with rough boundary conditions.  Surprisingly, there
barely exists any differences comparing to Fig.~\ref{msdxy}(b),
which depicts the MSDs of the fourth layer.  The particles overall
show aging behavior with slower dynamics for larger $t_{\rm age}$.
In contrast to the rough boundary, the MSD curves for the first
layer next to the smooth boundary look strikingly different
from the bulk case, as seen by comparing Fig.~\ref{msdxy}(c)
and (d).  In all four time groups the MSD curves in the first
layer are slightly smaller than those in the fourth layer. The
smooth wall greatly restricts particle mobility, similar to
what has been seen for dense colloidal liquids near smooth
walls.\cite{edmond12}  Moreover, unlike the fourth layer, where
the MSD curves strictly follow the aging order, the aging process
seems to reach a $t_{\rm age}$-independent state by $t_{\rm age}
= 8$~min.  This is likely because the dynamics in this layer are
extremely slow, including the aging dynamics.  This explanation
is also consistent with the pronounced first layer density peak
seen in Fig.~\ref{numdensity}(a).

\begin{figure}
\centering
\includegraphics[width=8cm,keepaspectratio]{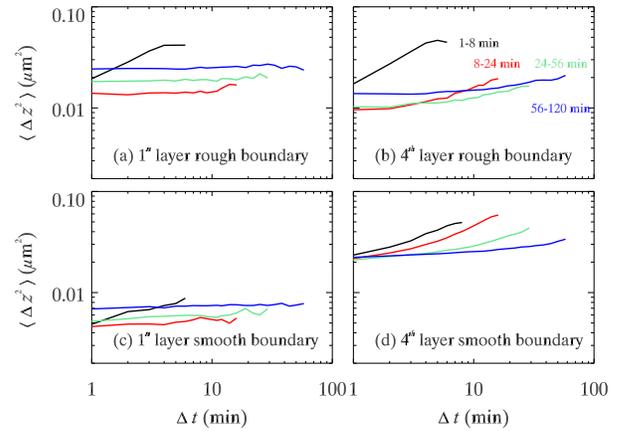}
\caption{The mean square displacement along the direction
perpendicular to the boundary ($z$)
calculated during four different $t_{\rm age}$ regimes as indicated.
Data are for (a)
$1^{st}$ layer with a rough boundary, (b) $4^{th}$ layer with
a rough boundary, (c) $1^{st}$  layer with a smooth boundary,
(d) $4^{th}$ layer with a smooth boundary.  
All displacements are also normalized by the
large particle diameter $d_L$.
}
\label{msdz} 
\end{figure}

Figure \ref{msdz} shows the MSD data for the $z$ component of motion,
perpendicular to the boundary.  The results are similar to the
MSD data of Fig.~\ref{msdxy}, with the exception that the layers
closest to the boundaries show less motion [panels (a,c)] for
both rough and smooth boundary conditions.  The increase in the
height of the MSD curves with age in Fig.~\ref{msdz}(a,c) is due
to photobleaching, but otherwise those MSD curves are fairly
flat.  Here the first layer for the rough boundary shows some
differences with the bulk behavior [compare panels (a) and (b)].
The contrast between first layer and bulk is stronger for the
smooth boundary condition [compare panels (c) and (d)].

\begin{figure}
\centering
\includegraphics[width=8cm,keepaspectratio]{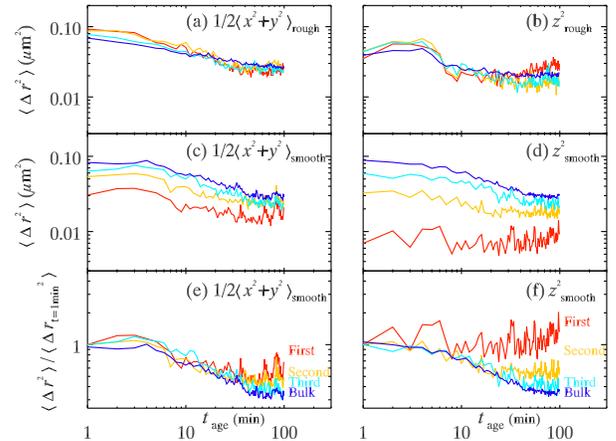}
\caption{Average distance particles move over $\Delta t= 20$~min,
as a function of aging time $t_{\rm age}$.
The curve colors indicate the layer number as labeled in
panels (e,f).  Panels show data for motions parallel to the
boundary ($1/2 \langle x^2 + y^2 \rangle$) and perpendicular to the
boundary ($\langle z^2 \rangle$) for rough and smooth boundaries
as indicated.  In panels (a-d) the data are normalized by the
large particle diameter $d_L$.  In panels (e,f) the data are
normalized by their initial values.
}
\label{dis} 
\end{figure}

To better understand the influence of the boundaries, we consider
a complementary analysis, examining $\langle \Delta r^2 \rangle$
at a fixed $\Delta t$ and varying $t_{\rm age}$.  We choose $\Delta
t=20$~min, where Figs.~\ref{msdxy},\ref{msdz} show that the particles' average
movement decreases with increasing $t_{\rm age}$ in both smooth
and rough boundaries.  Figure \ref{dis} shows the data divided by
rough boundary condition (panels a,b) and smooth boundary
condition (panels c,d), for motion parallel and perpendicular to
the boundaries (left and right panels respectively).  The colors
indicate different layers, as labeled in panels (e,f).  
The overall decreasing trend of all the curves with larger
$t_{\rm age}$ is the signature of aging, with the logarithmic
$t_{\rm age}$ axis making apparent that the rate of decrease
itself is slower in older samples.  The data suggest the sample
is still aging at the longest times observed in our experiment,
although even reaching a state-steady for $\Delta
t=20$~min does not preclude the sample from still having an aging
signal at longer $\Delta t$.

For the rough boundaries [Fig.~\ref{dis}(a,b)], the data collapse
for all layers confirming that the boundary appears to have a
negligible influence on the dynamics.  However, for the smooth
boundary condition, the wall-induced structures bring significant
differences for motion parallel to the boundary [Fig.~\ref{dis}(c)]
and even larger differences in the perpendicular direction 
[Fig.~\ref{dis}(d)].  Both types of motion are slower
closer to the wall.  For the motion perpendicular to the boundary
(panel d), the motion in the first layer is around ten
times smaller than the bulk.  Moreover, unlike other layers, we
do not observe an aging signal in the first layer -- the curve is
essentially flat.  The lack of observed aging behavior of $\Delta
z^2$ suggests that this
first layer has very slow dynamics.
Of course, the perpendicular motion
in the first layer is bounded at $z=0$, but the displacements we
observe are much smaller than for the first layer next to the rough
wall, which has a similar constraint on perpendicular motion.
Our observations of nearly immobile particles with
no aging signature in this first layer matches results from thin
polymer films near attractive silica substrates.\cite{priestley05}

As a different way of understanding how the aging process
changes near the smooth boundary, we normalize $\langle \Delta
r^2(t)\rangle$ by $\langle \Delta r^2(t_{\rm age}=1$~min$)
\rangle$ as shown in Fig~\ref{dis}(e,f).  For both motion parallel
and perpendicular to the boundary, the data collapse moderately
well for $t_{\rm age} \lesssim 10$~min, indicating an initial
aging trend.  For $t_{\rm age} \gtrsim 10$~min, the first and
second layers nearly stop evolving while the other layers
are still aging.  This is especially true for the $z$ motion
(panel f).

\begin{figure}
\centering
\includegraphics[width=8cm,keepaspectratio]{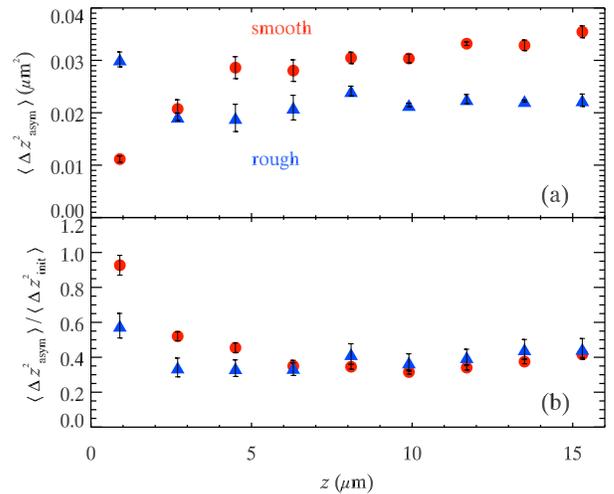}
\caption{(a) The $t_{\rm age}$ average of $\Delta z^2$
for the last 15~min of Fig.~\ref{dis}(b,d) plotted 
as a function of $z$, with the average done over all particles in
a layer (as defined in Fig.~\ref{numdensity}).  The $z$ value
is the center of each layer over which the average is taken.
(b) The same data normalized by the mean value of $\Delta z^2$
for $t_{\rm age} \leq 5$~min.  This represents the slowing seen
due to aging; the data close to 1.0 show little or no aging
behavior.  The error bars represent the variability in the
results when different ranges for the $t_{\rm age}$-averaging
are used.
}
\label{binz} 
\end{figure}

To further explore the relation between the layering structures
and motion perpendicular to the boundary we define
$\langle \Delta z^2_{\rm asym} \rangle$.  This is the average of
the data of Fig.~\ref{dis}(b,d) in the asymptotic regime, that
is, for $t_{\rm age} \geq 85$~min.  The
results are plotted as a function of the distance from the wall 
in Fig.~\ref{binz}(a).  The smooth data (red
circles) smoothly increases as $z$ increases.  The rough data
(blue triangles) are fairly constant, with the exception of the
first layer ($z = 0.8$~$\mu$m) which is larger.  As argued above
based on the flatness of the data in Fig.~\ref{msdz}(a),
this increase in the first layer is likely due to photobleaching
than true motion.  For $z > 5$~$\mu$m the differences between
smooth and rough data are likely due to image quality which
artificially increases the MSD.\cite{poon12}  To
account for this, in Fig.~\ref{binz}(b) are normalized by the
value of $\langle \Delta z^2 \rangle$ averaged over $t_{\rm age}
\leq 5$~min.  This collapses the data for $z > 5$~$\mu$m.  These
data are related to the amount the dynamics slow as the
sample ages, with 1 corresponding to no slowing and smaller
values indicating slowing with age.  
The value close to 1 for the smooth boundary
condition indicates that the first layer barely ages, consistent
with the similarity of the MSD data of Fig.~\ref{msdz}(c) and the
horizontal red line in Fig.~\ref{dis}(d).  The decrease in the
data of Fig.~\ref{binz}(b) as $z$ increases shows a return to the
normal aging seen in the bulk.

\begin{figure}
\centering
\includegraphics[width=7cm,keepaspectratio]{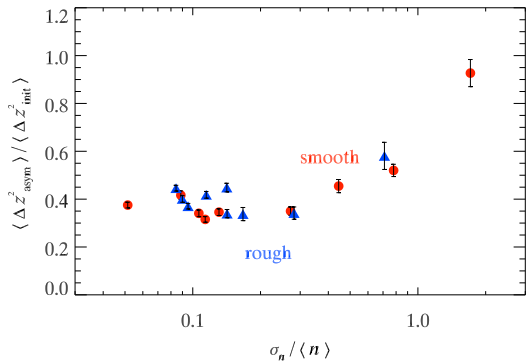}
\caption{The large $t_{\rm age}$ motion in $z$ 
plotted as a function of the standard deviation of number
density $\sigma_{\rm n}$ over the mean number density $\langle
n \rangle$ where these quantities are defined within each layer
(see Fig ~\ref{numdensity}).  The data for the vertical
coordinate correspond to that of Fig.~\ref{binz}(b).
}
\label{sigma} 
\end{figure}

The qualitative similarity of the rough and smooth data in
Fig.~\ref{binz}(b) motivate an attempt to collapse the data
by a horizontal shift.  Noting that the number density data
of Fig.~\ref{numdensity}(b) look like a horizontally shifted
portion of the data of Fig.~\ref{numdensity}(a), we use the local
layer structure as a possible way to explain the dynamical data.
We quantify the structure using the standard deviation of $n(z)$
within a layer divided by its mean.   This nearly collapses the
data (to within fluctuations of $\sim 20\%$) accounting for most of
the effect.  The data for $\sigma_n/\langle n \rangle \lesssim 0.2$
are essentially the bulk region.  Thus 
the difference in dynamics between
the smooth and rough boundaries we observe can be explained by
the difference in particle layering that occurs next to these
two interfaces.

\section{Conclusions}
\label{sectionConclusions}

In our experiment we study aging by observing particle motion
in a colloidal glass near smooth and rough boundaries.  Both
samples exhibit aging in their bulk.  Near a smooth boundary,
the particles form layers against the boundary such that in the
two layers closest to the wall, motion is greatly diminished.
For a smooth wall, we observe the influence of the boundary extends
up to $\approx 6$ layers ($\approx 4$ large particle diameters)
into the sample.  The observations of a gradient in dynamics near
the smooth wall are qualitatively similar to prior observations of
gradients near interfaces in glassy materials.  Direct evidence
for gradients in dynamics has been seen in molecular dynamics
simulations \cite{kob00,scheidler02,scheidler04} and colloidal
experiments.\cite{sarangapani11,hunter14}  In other experiments
the influences of the boundaries are inferred from local probes
near the boundary (e.g., Ref.~\cite{priestley05}) or fitting the
data to models assuming boundary effects (e.g., Ref.~\cite{pye10}).

Here we not only see the gradient in dynamics, but observe that
this gradient in dynamics is directly related to a gradient in the
structural properties.  For a rough boundary, the wall-induced
structure is greatly reduced and the dynamics appear more
bulk-like near the boundary, being similar to that far into
the bulk.  By comparing the local dynamics near the rough and
smooth boundaries, our results suggest that the dominant factor
modifying aging dynamics near a boundary is the structure caused
by the presence of the boundary.  By presenting a rough amorphous
boundary, the structure is more bulk-like and thus the dynamics
are more bulk-like.  A fruitful area for future work would be to
explore boundary textures that have intermediate influences on
layering structure.

These experimental results on colloidal glasses suggest a viable
means by which neutral rough amorphous boundaries may be
implemented in computer simulations.  This is an issue that
computational studies on the influence of interfacial effects on
local dynamics have been struggling with for more than two
decades,\cite{baschnagel05,kob00,scheidler02,scheidler04,hanakata15,davris15,fehr95}
and has relevance for the
implementation of theoretical point-to-set
studies.\cite{biroli08,li14,hocky14,himanagamanasa15}  The
method employed in the present study creates a rough amorphous
boundary by randomly sticking particles to a smooth wall at
approximately 30-50\% surface coverage.  The local aging dynamics
we observe near such a rough boundary appear nearly bulk-like
with little deviation from bulk particle densities.

This work was supported by the National Science Foundation
(DMR-1609763).

%\bibliography{congpaper}
\bibliography{eric}

\end{document}